\documentclass[fleqn,usenatbib]{mnras}
\usepackage{newtxtext,newtxmath}
\usepackage[T1]{fontenc}

\DeclareRobustCommand{\VAN}[3]{#2}
\let\VANthebibliography\thebibliography
\def\thebibliography{\DeclareRobustCommand{\VAN}[3]{##3}\VANthebibliography}

\usepackage{graphicx}	
\usepackage{amsmath}	
\usepackage{multirow}
\usepackage{soul}
\usepackage[shortcuts]{extdash}
\usepackage{orcidlink}
\usepackage[normalem]{ulem}
\usepackage{fixfoot}
\usepackage{eso-pic}

\DeclareFixedFootnote{\repnote}{Available at \url{https://github.com/des-science/DES-SN5YR} along with the statistical+systematic covariance matrix used in this analysis.} 
\newcommand{\snbao}{DES-SN5YR + DESI-BAO}
\newcommand{\bestfit}{$H_0=67.19^{+0.66}_{-0.64}\mathrm{~km} \mathrm{~s}^{-1} \mathrm{~Mpc}^{-1}$}
\newcommand{\kmsMpc}{km\,s$^{-1}$\,Mpc$^{-1}$}
\newcommand{\mycite}[2]{(\citeauthor{#1} \citeyear{#1}, hereafter #2)}
\newcommand{\mycitee}[2]{(#2;~\citeauthor{#1} \citeyear{#1})}

\title[An updated measurement of the Hubble constant using the Inverse Distance Ladder]{The Dark Energy Survey Supernova Program: An updated measurement of the Hubble constant using the Inverse Distance Ladder}

\author[R. Camilleri et al.]{
\parbox{\textwidth}{
R.~Camilleri,$^{1}$\thanks{E-mail: uqrcamil@uq.edu.au (RC)}
T.~M.~Davis,$^{1}$
S.~R.~Hinton,$^{1}$
P.~Armstrong,$^{2}$
D.~Brout,$^{3}$
L.~Galbany,$^{4,5}$
K.~Glazebrook,$^{6}$
J.~Lee,$^{7}$
C.~Lidman,$^{8,2}$
R.~C.~Nichol,$^{9}$
M.~Sako,$^{7}$
D.~Scolnic,$^{10}$
P.~Shah,$^{11}$
M.~Smith,$^{12}$
M.~Sullivan,$^{12}$
B.~O.~S\'anchez,$^{10,13}$
M.~Vincenzi,$^{10}$
P.~Wiseman,$^{12}$
S.~Allam,$^{14}$
T.~M.~C.~Abbott,$^{15}$
M.~Aguena,$^{16}$
F.~Andrade-Oliveira,$^{17}$
J.~Asorey,$^{18}$
S.~Avila,$^{19}$
D.~Bacon,$^{20}$
K.~Bechtol,$^{21}$
S.~Bocquet,$^{22}$
D.~Brooks,$^{11}$
E.~Buckley-Geer,$^{23,14}$
D.~L.~Burke,$^{24,25}$
A.~Carnero~Rosell,$^{26,16}$
D.~Carollo,$^{27}$
J.~Carretero,$^{19}$
F.~J.~Castander,$^{4,5}$
C.~Conselice,$^{28,29}$
L.~N.~da Costa,$^{16}$
M.~E.~S.~Pereira,$^{30}$
S.~Desai,$^{31}$
H.~T.~Diehl,$^{14}$
S.~Everett,$^{32}$
I.~Ferrero,$^{33}$
B.~Flaugher,$^{14}$
J.~Frieman,$^{14,34}$
J.~Garc\'ia-Bellido,$^{35}$
E.~Gaztanaga,$^{4,20,5}$
G.~Giannini,$^{19,34}$
R.~A.~Gruendl,$^{36,37}$
K.~Herner,$^{14}$
D.~L.~Hollowood,$^{38}$
K.~Honscheid,$^{39,40}$
D.~Huterer,$^{17}$
D.~J.~James,$^{3}$
S.~Kent,$^{14,34}$
K.~Kuehn,$^{41,42}$
O.~Lahav,$^{11}$
S.~Lee,$^{32}$
G.~F.~Lewis,$^{43}$
M.~Lima,$^{44,16}$
J.~L.~Marshall,$^{45}$
J. Mena-Fern{\'a}ndez,$^{46}$
R.~Miquel,$^{47,19}$
J.~Myles,$^{48}$
R.~L.~C.~Ogando,$^{49}$
A.~Palmese,$^{50}$
A.~Pieres,$^{16,49}$
A.~A.~Plazas~Malag\'on,$^{24,25}$
A.~K.~Romer,$^{51}$
A.~Roodman,$^{24,25}$
S.~Samuroff,$^{52}$
E.~Sanchez,$^{53}$
D.~Sanchez Cid,$^{53}$
M.~Schubnell,$^{17}$
I.~Sevilla-Noarbe,$^{53}$
E.~Suchyta,$^{54}$
N.~Suntzeff,$^{45}$
M.~E.~C.~Swanson,$^{36}$
G.~Tarle,$^{17}$
B.~E.~Tucker,$^{2}$
A.~R.~Walker,$^{15}$
and N.~Weaverdyck$^{55,56}$ (DES Collaboration)
}
\vspace{0.4cm}
\\
\textit{Affiliations are listed at the end of the paper}
}

\date{Accepted XXX. Received YYY; in original form ZZZ}

\AddToShipoutPictureBG*{%
  \AtPageUpperLeft{%
    \hspace{0.75\paperwidth}%
    \raisebox{-1.5\baselineskip}{%
      \makebox[0pt][l]{\textnormal{DES-2024-835}}}
}}%

\AddToShipoutPictureBG*{%
  \AtPageUpperLeft{%
    \hspace{0.75\paperwidth}%
    \raisebox{-2.5\baselineskip}{%
      \makebox[0pt][l]{\textnormal{FERMILAB-PUB-24-0290-PPD}}
}}}%

\pubyear{\the\year{}}

\begin{document}
\label{firstpage}
\pagerange{\pageref{firstpage}--\pageref{lastpage}}
\maketitle

\begin{abstract}
We measure the current expansion rate of the Universe, Hubble's constant $H_0$, by calibrating the absolute magnitudes of supernovae to distances measured by Baryon Acoustic Oscillations.  This `inverse distance ladder' technique provides an alternative to calibrating supernovae using nearby absolute distance measurements, replacing the calibration with a high-redshift anchor.  We use the recent release of 1829 supernovae from the Dark Energy Survey spanning $0.01<z<1.13$ anchored to the recent Baryon Acoustic Oscillation measurements from DESI spanning $0.30 <z_{\mathrm{eff}}<2.33$. To trace cosmology to $z=0$, we use the third, fourth and fifth-order cosmographic models, which, by design, are agnostic about the energy content and expansion history of the universe. With the inclusion of the higher-redshift DESI-BAO data, the third-order model is a poor fit to both data sets, with the fourth-order model being preferred by the Akaike Information Criterion. Using the fourth-order cosmographic model, we find \bestfit, in agreement with the value found by Planck without the need to assume Flat-$\Lambda$CDM. However the best-fitting expansion history differs from that of Planck, providing continued motivation to investigate these tensions. 
\end{abstract}



\section{Introduction} \label{sec:intro}
Resolving the tension between late-time and early-time measurements of the Hubble constant ($H_0$) is one of the most significant challenges presented by the standard cosmological model. The Planck Collaboration  \citep[hereafter Planck;][]{2020_planck}, which measures the Cosmic Microwave Background (CMB) radiation, estimates the local expansion rate to be $H_{0}=67.4 \pm 0.5 \mathrm{~km} \mathrm{~s}^{-1} \mathrm{~Mpc}^{-1}$ assuming a spatially flat $\Lambda$CDM universe. This value is in $\sim5\sigma$ tension and substantially lower than that determined by the SH0ES collaboration \citep{SH0ES_2021}, $H_{0}=73.04 \pm 1.04 \mathrm{~km} \mathrm{~s}^{-1} \mathrm{~Mpc}^{-1}$ using a distance ladder consisting of Cepheid calibrated supernova (SN) luminosity distances from Pantheon+ \citep{scolnic2021pantheon,2022_pantheon_analysis}, assuming only the cosmological principle, that is our universe is homogeneous and isotropic at large scales. For a recent review of the Hubble constant, see \citet{Shah_2021}.

The discrepancy between early and late-time measurements might imply new physics and has resulted in alternate cosmological models being proposed. Recent investigations into alternate models have shown many to be compatible with various data sets \citep{Zhang_2017, Dam_2017, camilleri24}. 
There also might be unaccounted for systematics or some unknown physical phenomenon that has not been taken into account in the current models of the universe's expansion and has motivated an increase in new, independent ways to determine $H_0$. 

Since SN Ia are relative distance indicators, their observed magnitude must be calibrated using an absolute distance measurement. In the distance ladder approach used by SH0ES, parallax measurements to nearby Cepheid variable stars are used to calibrate the luminosities of SNe Ia. Therefore, one alternate way to determine $H_0$ involves replacing Cepheids as calibrators for SNe Ia, avoiding unknown systematics associated with this specific method. \citet{Freedman_2021} and \citet{Anand_2022} calibrate the SN Ia with tip of the red giant branch (TRGB) distances to host galaxies, and found $H_0 = 69.8\pm0.6~(\mathrm{stat})\pm1.6~(\mathrm{sys}) \mathrm{~km} \mathrm{~s}^{-1} \mathrm{~Mpc}^{-1}$ and $H_0 = 71.5\pm1.8 \mathrm{~km} \mathrm{~s}^{-1} \mathrm{~Mpc}^{-1}$ respectively, both consistent to within $2\sigma$ of the SH0ES result. Recently, \cite{Scolnic_2023} also calibrated the SNe Ia with TRGB distances and measured a higher value finding $H_0=73.22 \pm 2.06\mathrm{~km} \mathrm{~s}^{-1} \mathrm{~Mpc}^{-1}$. For more details and a review on recent TRGB measurements see \cite{li2024tip}. Surface brightness fluctuations (SBF) of a host galaxy can also be used to calibrate SNe Ia \citep{Tonry1988, Blakeslee1999,Biscardi_2008, Blakeslee2009}. This technique was first used by \citet{Khetan2021}, who found $H_0 = 70.50\pm2.37~(\mathrm{stat})\pm3.38~(\mathrm{sys}) \mathrm{~km} \mathrm{~s}^{-1} \mathrm{~Mpc}^{-1}$, consistent with the value obtained by SH0ES and Planck. A later measurement by \citet{Garnavich_2023} found a higher value of $H_0 = 74.6\pm0.9~(\mathrm{stat})\pm2.7~(\mathrm{sys}) \mathrm{~km} \mathrm{~s}^{-1} \mathrm{~Mpc}^{-1}$. \cite{blake11} also show that SN can be used with the Alcock-Paczynski test to measure the cosmic history in a model independent way and determined $H(z) /\left[H_0(1+z)\right]$ in four different redshift slices to $10-15\%$ accuracy.

The approach we take here, is to use the inverse distance ladder method where the observed magnitudes of SNe Ia are calibrated using distance measurements from Baryon Acoustic Oscillations (BAOs) with a prior on the CMB sound horizon at the time of photon-baryonic decoupling after recombination. We then use the cosmographic approach \citep{Aubourg_2015, Macaulay_2019}, which is a smooth Taylor expansion of $H(z)$ (see Section~\ref{cosmography}) to theoretically trace the cosmology to $z=0$ assuming a FLRW metric and the Etherington distance duality relation\footnote{The Etherington distance duality relates the luminosity distance $D_L(z)$ to the angular diameter distance $D_A(z)$ by $D_L(z) = D_A(z)(1 + z)^2$. } \citep{etherington33}. 

This procedure was first used by \cite{Aubourg_2015}, who found $H_0=67.3\pm 1.1 \mathrm{~km} \mathrm{~s}^{-1} \mathrm{Mpc}^{-1}$ in agreement with the Planck result using 740 SNe Ia from the Joint Light Curve analysis \mycite{2014JLA}{JLA} and BAO measurements from the Baryon Oscillation Spectroscopic Survey (BOSS) Data Release Eleven \citep[DR11;][]{bossdr11}. 

Harnessing the same technique, \citet{Macaulay_2019} found $H_0 = 67.8\pm 1.3 \mathrm{~km} \mathrm{~s}^{-1} \mathrm{~Mpc}^{-1}$ using 207 SN Ia from the Dark Energy Survey (DES) Collaboration's \citep{Abbott_2019} 3-year SN release, with BAO measurements taken from \cite{Carter_2018} and the Baryon Oscillation Spectroscopic Survey Data Release Twelve \mycitee{Alam_2017}{BOSS DR12}. 

Recently, DES has released the largest single sample of SNe to date. The release consisted of 1829 SNe in the redshift range $0.10 < z < 1.13$ \citep{DES-SN5YR}. Furthermore, the Dark Energy Spectroscopic Instrument (DESI) team \citep{desicollaboration2024desi} has provided BAO measurements in seven redshift bins from over 6 million extragalactic objects in the redshift range $0.1< z < 4.2$. In this paper, we use these data sets from both DES and DESI to provide an updated measurement of $H_0$ using the inverse distance ladder technique.

This paper is organized as follows, in Section~\ref{sec:method&data} we provide the basic theory, methodology and data sets used in our analysis. We present our results in Section~\ref{sec:results} and conclude in Section~\ref{sec:conclusion}.

\section{Methodology \& Data}\label{sec:method&data}
\subsection{Cosmographic Expansion}\label{cosmography}
To trace the cosmology to $z=0$, we use the well-established cosmographic expansion, which is a smooth Taylor expansion of the scale factor $a$ that makes minimal assumptions about the underlying cosmological model but retains the assumptions of homogeneity and isotropy \citep{Visser_2004, Zhang_2017}. Here, we note the definition of the deceleration parameter,
\begin{equation}\label{eq:q_param}
    q = -\frac{1}{H^2}\frac{1}{a}\frac{d^2a}{dt^2},
\end{equation}
the jerk parameter,
\begin{equation}\label{eq:j_param}
    j = \frac{1}{H^3}\frac{1}{a}\frac{d^3a}{dt^3},
\end{equation}
the snap parameter, 
\begin{equation}\label{eq:s_param}
    s = \frac{1}{H^4}\frac{1}{a}\frac{d^4a}{dt^4},
\end{equation}
and the lerk parameter,
\begin{equation}\label{eq:l_param}
    l = \frac{1}{H^5}\frac{1}{a}\frac{d^5a}{dt^5}.
\end{equation}
With these definitions, the Hubble parameter can be expressed to fifth-order as,
\begin{align}
H(z)&=H_0\left[1+ \mathcal{H}_{1}z+ \mathcal{H}_{2}z^2+\mathcal{H}_{3}z^3+\mathcal{H}_{4}z^4\right] \label{hz}
\end{align}
where
\begin{flalign}
\mathcal{H}_{1} &= \left(1+q_0\right), && \nonumber \\ \nonumber
\mathcal{H}_{2} &= \frac{1}{2} \left(j_0-q_0^2\right), && \\ \nonumber
\mathcal{H}_{3} &= \frac{1}{6} \left( 3q_0^{2} + 3q_0^3 - 4q_0 j_0 - 3 j_0 - s_0\right), &&\\ \nonumber
\mathcal{H}_{4} &= \frac{1}{24} (-12q^2_0 - 24 q^3_0 - 15q^4_0 + 32q_0j_0 + 25q^2_0j_0 &&\\ \nonumber
&+ 7q_0s_0 + 12j_0 - 4j^2_0+8s_0+l_0),
\end{flalign}
and $q_0$, $j_0$ $s_0$ and $l_0$ are the current epoch deceleration, jerk, snap and lerk parameters respectively. The luminosity distance, $D_L(z)$ for a spatially flat universe is given by,
\begin{align}
D_L(z)&=z+\mathcal{D}_1 z^2+\mathcal{D}_2 z^3+\mathcal{D}_3 z^4+\mathcal{D}_4z^5 \label{lumdist}
\end{align}
where,
\begin{flalign}
\mathcal{D}_{1} &= \frac{1}{2} \left(1-q_0\right), \nonumber && \\
\mathcal{D}_{2} &= -\frac{1}{6} \left(1-q_0 - 3q_0^2 + j_0\right), \nonumber&& \\
\mathcal{D}_{3} &= \frac{1}{24} \left(2-2q_0-15q^2_0 + 5j_0+10q_0 j_0 + s_0 \right), \nonumber&&\\
\mathcal{D}_{4} &= \frac{1}{120} (-6 +6q_0 + 81q^2_0 + 165q^3_0 + 105q^4_0 + 10j^2_0 \nonumber && \\ \nonumber
&-27j_0 - 110q_0j_0-105q^2_0j_0-15q_0s_0-11s_0-l_0).
\end{flalign}

\subsection{DES-SN5YR}
In this analysis, we use the DES-SN5YR sample containing 1829 likely SNe Ia. The DES-SN5YR sample is the largest and deepest single sample survey to date consisting of 1635 SNe ranging in redshift from $0.10$ to $1.13$ from the the full DES survey and is complemented by 194 spectroscopically confirmed low-redshift SN Ia. For more details see \citet{vincenzi24} and \citet{DESDATA24}. 

The DES-SN5YR Hubble diagram\repnote\ includes corrected SN apparent magnitudes, distance moduli (calculated using an assumed $H_0$ and absolute magnitude, $M_B$), heliocentric redshifts, CMB corrected redshifts and redshifts with both CMB and peculiar velocity corrections. The DES Hubble diagram is not anchored to a specific calibrator and therefore on its own offers little information regarding $H_0$ or the SN Ia peak absolute magnitude, $M_B$ due to a degeneracy between these parameters. However, this degeneracy can be broken with BAO data and therefore we use the \textit{corrected} apparent magnitudes provided by DES and calculate our data vector as,
\begin{equation}
    \mu_{\mathrm{data}}(M_B) = m_{b,\mathrm{data}} - M_B
\end{equation}
where $M_B$ is a free parameter. The distance moduli can then be calculated as,
\begin{equation}
    \mu_{\mathrm{theory}}(z, \Theta) = 5 \log_{10}[D_L(z, \Theta)/1~\mathrm{Mpc}] + 25
\end{equation}
where $D_L(z, \Theta)$ is given in equation~(\ref{lumdist}) and depends on the set of cosmological parameters, $\Theta$. In this analysis, we fit cosmographic expansion three ways: to third-order (equation~\ref{lumdist} excluding the $z^4$ and $z^5$ terms), fourth-order (equation~\ref{lumdist} excluding the $z^5$ term) and fifth-order (equation~\ref{lumdist}) with $\Theta = \{H_0, q_0, j_0\}$, $\Theta = \{H_0, q_0, j_0, s_0\}$ and $\Theta = \{H_0, q_0, j_0, s_0, l_0\}$ respectively.

We compute the difference between data and theory for every $i$th SN, $D_i(M_B, \Theta) = \mu_{\mathrm{data}}(M_B) - \mu_{\mathrm{theory}}(z, \Theta)$, and find the minimum of
\begin{equation}\label{eq:chi2_sn}
\chi^{2}_{\mathrm{SN}}(M_B, \Theta)=\vec{D}^{T}\mathcal{C}^{-1}_{\mathrm{SN}} \vec{D},
\end{equation}
where $\mathcal{C}^{-1}_{\mathrm{SN}}$ is the inverse covariance matrix inlcuding both statistical and systematic errors.\repnote\

\subsection{DESI-BAO}
The sound horizon at the time of baryon decoupling in the early universe, i.e., the drag epoch (at $z \simeq 1060$), left an imprint in the distribution of matter, which is detectable in the galaxy distribution. These BAO serve as cosmological standard rulers \citep{Blake_2003, Seo_2003, Linder_2003, McDonald_2007, Alam_2017, desicollaboration2024desi}. In this work, we use measurements of the BAO provided by the DESI collaboration \citep{desicollaboration2024desi}. The DESI-BAO provide twelve measurements in seven redshift bins from over 6 million extragalactic objects in the redshift range $0.1< z < 4.2$.

BAO measurements have a physical scale set by the sound horizon, $r_s$ at the end of the drag epoch, $r_d \equiv r_s(z_*)$, and are observed from pairs of galaxies averaged over all angles. Measurements are generally quoted as $D_H(z)/r_d$ for separation vectors of the pairs that are oriented parallel to the line of sight and $D_M(z)/r_d$ for pairs oriented perpendicular to the line of sight with both results being correlated. For certain redshift bins with low signal-to-noise ratio, the volume-averaged quantity $D_V(z)/r_d$ is quoted. The distances $D_H(z), D_M(z)$ are the Hubble and transverse comoving distances respectively and can be calculated as,
\begin{align}
D_H(z, \Theta)&=c/H(z, \Theta), \\
\intertext{where $H(z, \Theta)$ is given in equation~(\ref{hz}) and}
D_M(z, \Theta)&=D_L(z, \Theta)/(1+z),
\intertext{where $D_L(z, \Theta)$ is given in equation~(\ref{lumdist}). The dilation scale, $D_V(z)$ is a combination of the two distances and defined as, }
D_V(z, \Theta) &\equiv\left[z D_M^2(z, \Theta) D_H(z, \Theta)\right]^{1 / 3}.
\end{align}
Lastly, we require knowledge of $r_d$, which depends on the baryon density and total matter density in the early universe.  In this work we adopt a Gaussian prior of $r_d\sim \mathcal{N}(147.46,\,0.28)\,$ Mpc determined using constraints from \citet{Lemos23}, who removed late-time cosmology dependence from the CMB likelihoods (Planck Flat-$\Lambda$CDM) associated with modelling the late-Integrated Sachs-Wolfe (late-ISW) effect, the optical depth to reionization, CMB lensing and foregrounds.

To constrain the DESI-BAO data we find the minimum of 
\begin{equation}\label{eq:chi2_bao}
\chi^{2}_{\mathrm{BAO}}(r_d, \Theta)=\vec{\Delta}^{T}\mathcal{C}^{-1}_{\mathrm{BAO}} \vec{\Delta},
\end{equation}
where $\Delta(r_d, \Theta)$ is the difference between the measurements and the associated values determined with the cosmographic model and $\mathcal{C}^{-1}_{\mathrm{BAO}}$ is the inverse covariance matrix provided with the DESI-BAO.

\begin{table*}[hp]
    \centering
    \caption{Best fit parameters determined with the combined \snbao~datasets. These results are the medians of the marginalised posterior with 68.27\% integrated uncertainties. The $\Delta$AIC with respect to the $4^{\mathrm{th}}$ order cosmographic model is quoted in the final column.}
    \label{tab:bestfit_params}
    \renewcommand{\arraystretch}{1.3}
    \begin{tabular}{ccccccccc}
        \hline
		Model & $H_0$ & $M_B$ & $\Omega_{\mathrm{m}}$ & $q_0$ & $j_0$ & $s_0$ & $l_0$ & $\Delta$AIC \\ 
		\hline
		Flat-$\Lambda$CDM & $67.67^{+0.69}_{-0.68}$ & $-19.394^{+0.017}_{-0.016}$ & $0.322\pm 0.012$ & -- & -- & -- & -- & 6.1  \\
		$3^{\mathrm{rd}}$ order & $70.03^{+0.66}_{-0.65}$ & $-19.344^{+0.016}_{-0.015}$ & - &  $-0.551\pm 0.025$ & $0.703^{+0.056}_{-0.057}$ & -- & -- & 89.9\\ 
		$4^{\mathrm{th}}$ order & $67.19^{+0.66}_{-0.64}$ & $-19.375^{+0.018}_{-0.017}$ & - & $-0.342\pm 0.013$ & $0.255^{+0.087}_{-0.091}$ & $-0.494^{+0.046}_{-0.048}$ & -- & 0.0 \\ 
        $5^{\mathrm{th}}$ order & $67.26^{+0.73}_{-0.71}$ & $-19.381\pm 0.018$ & - & $-0.391^{+0.032}_{-0.033}$ & $0.539^{+0.031}_{-0.030}$ & $-0.45^{+0.32}_{-0.30}$ & $2.6^{+1.1}_{-1.2}$ & 2.3\\  
		\hline
    \end{tabular}
\end{table*}
\begin{figure*}[hp]
    \centering
    \includegraphics[width=0.95\linewidth]{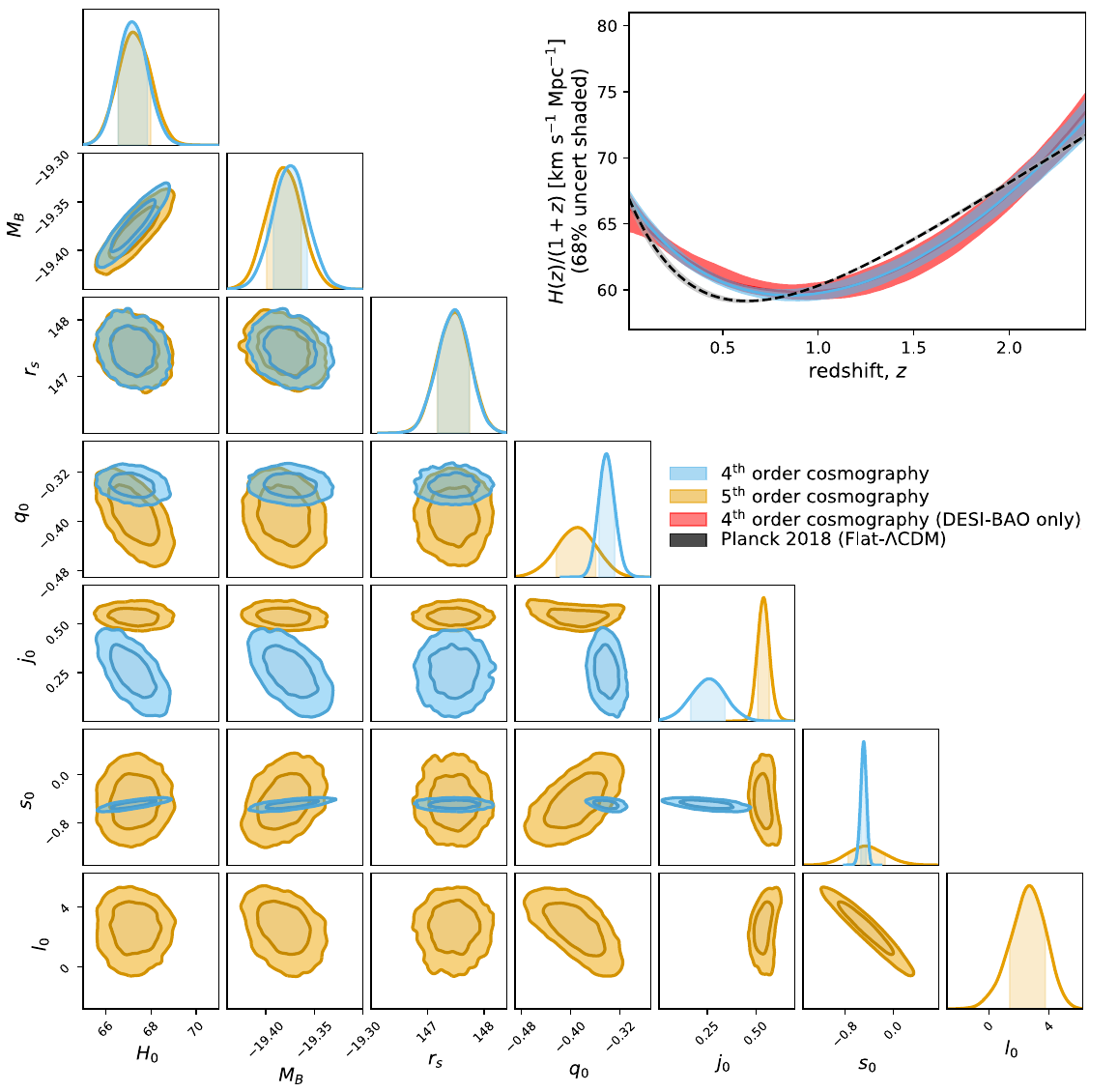}
    \caption{The constraints on $H_0$, $M_B$, $r_s$ and the $4^{\mathrm{th}}$ (blue) and $5^{\mathrm{th}}$ (orange) order cosmographic parameters using the combined \snbao~ datasets. Contours represent the 68.3\% and 95.5\% confidence intervals. The median of the marginalised posterior and cumulative 68.27\% confidence intervals are given in Table~\ref{tab:bestfit_params}. In the top right of the figure, we also present the best fit expansion history for the $4^{\mathrm{th}}$ order cosmographic model (blue) determined with the \snbao~datasets. For comparison, we include the expansion history from Planck, which was determined by analysing the CMB and assuming Flat-$\Lambda$CDM (black dashed line) and the  $4^{\mathrm{th}}$ order cosmographic model fit to the DESI-BAO alone (red).  }
    \label{fig:H0_constraint}
\end{figure*}

\subsection{Fitting for \texorpdfstring{$H_0$}{H0}}
To determine $H_0$ using the inverse distance ladder technique we perform a combined fit to the DES-SN5YR and DESI-BAO, breaking the degeneracy between $M_B$ and $H_0$. The likelihoods are combined as,
\begin{equation}\label{eq:chi2_tot}
\chi^{2}_{\mathrm{tot}}(M_B, r_d, \Theta)=\chi^{2}_{\mathrm{BAO}}(r_d, \Theta) + \chi^{2}_{\mathrm{SN}}(M_B, \Theta)
\end{equation}
and minimised using the dynamic nested sampling package, \texttt{Dynesty} \citep{dyn3, dyn4, dyn5, dyn1,dyn2} with 500 live points. We also used \texttt{Dynesty}'s inbuilt stopping function, which decides when to stop sampling based on a posterior and evidence error threshold, with the default values.

\section{Results \& Discussion}\label{sec:results}

\begin{figure}
    \centering \includegraphics{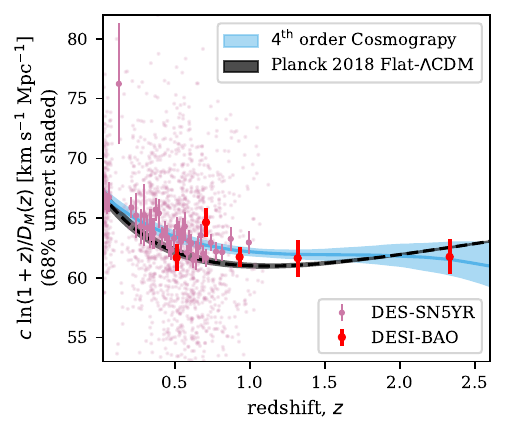}
    \caption{Illustration of the inverse distance ladder method. The pink supernova data points from DES-SN5YR have been calibrated to the red data points from DESI-BAO, and the resulting y-intercept gives $H_0$. For DESI-BAO we plot only the five $D_M(z)/r_d$ points and their statistical uncertainties.  For the DES-SN5YR sample we show both the individual SN events (transparent) and the redshift-binned SN distance moduli (opaque with redshift-binned statistical uncertainties) after calibration. The blue line represents the best fit $4^{\mathrm{th}}$ order cosmographic model determined with the \snbao~datasets. The black dashed line shows the best fit from Planck \citep{2020_planck}, determined by analysing the CMB and assuming Flat-$\Lambda$CDM. }
    \label{fig:dm_expansion}
\end{figure}

Best fit parameters determined with the combined \snbao~datasets are shown in Table~\ref{tab:bestfit_params} and Fig.~\ref{fig:H0_constraint}. We plot the best fit expansion histories in the top right of Fig.~\ref{fig:H0_constraint} and Fig.~\ref{fig:dm_expansion}. These results are the medians of the marginalised posterior with 68.27\% integrated uncertainties determined with the `cumulative' option in the Python package \texttt{ChainConsumer} \citep{hinton16}.

To assess whether the additional parameters used in the higher order cosmographic models are required given the data we use the Akaike Information Criterion ${\rm AIC}\equiv 2k-2\ln\mathcal{L}^{\max}$ \citep{1100705}, where $k$ is the number of parameters in the model. In the final column of Table~\ref{tab:bestfit_params} we quote the $\Delta$AIC relative to the $4^{\mathrm{th}}$ order cosmographic model. To assess the strength for or against a model, \citet{Trotta08} suggests that $\Delta>2$, $\Delta>5$ and $\Delta>10$ indicates weak, moderate and strong evidence respectively, against the model with the higher $\Delta$ value. 

We find strong evidence against the $3^{\mathrm{rd}}$ order cosmographic model with $\Delta$AIC $=89.9$ that is driven by the model's poor fit to both data sets with a $\Delta \chi^2 =81$ relative to the $4^{\mathrm{th}}$ order cosmographic model. For reference, we include the Flat-$\Lambda$CDM fit. This result shows that the $4^{\mathrm{th}}$ order model is a good fit to the data and is moderately preferred over Flat-$\Lambda$CDM ($\Delta$AIC $=6.1$). Note that the $3^{\mathrm{rd}}$ order cosmographic model has been shown to be a good fit to the DES-SN5YR alone \citep[although $4^{\mathrm{th}}$ order is still preferred by AIC; see][]{camilleri24}. However, with the inclusion of the higher-redshift DESI data, a more flexible model is needed.  We therefore do not discuss the $3^{\rm rd}$ order model any further. Both of our key results, presented below, are quoted using the $4^{\mathrm{th}}$ order cosmographic expansion. However, we note that there is no preference for or against the $4^{\mathrm{th}}$ and $5^{\mathrm{th}}$ order expansions based on the AIC. This choice does not impact the conclusions of this paper with both models giving consistent constraints on $H_0$.

\begin{figure}
    \centering \includegraphics[width=\linewidth]{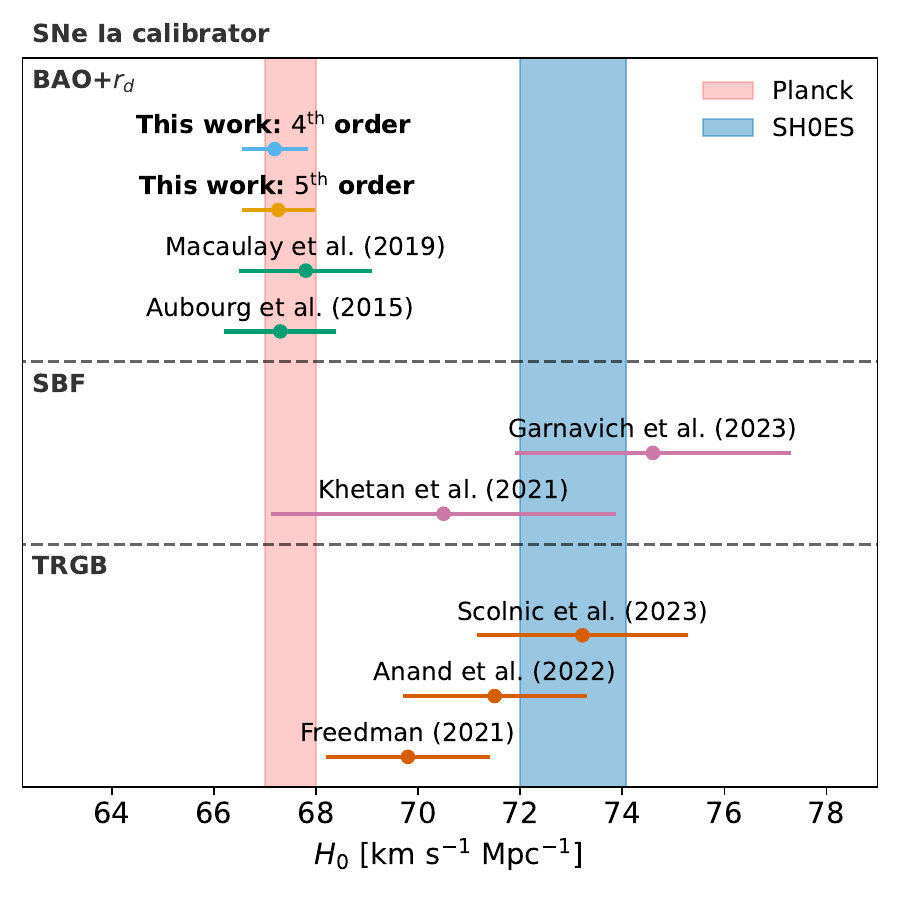}
    \caption{A comparison of Hubble constant values using different SNe Ia calibrators and the results from this work for the $4^{\mathrm{th}}$ (blue) and $5^{\mathrm{th}}$ (orange) order cosmographic fit. The red shaded regions represent the 68\% confidence interval for the Planck (red) and SH0ES (blue) result.}
    \label{fig:h0_comparison}
\end{figure}

The combined \snbao~datasets yield \bestfit~(median of the marginalised posterior with 68.27\% integrated uncertainties) when using the $4^{\mathrm{th}}$ order cosmographic model. This measurement is consistent with the Planck+Flat-$\Lambda$CDM measurement of $H_0=67.4\pm0.5\mathrm{~km} \mathrm{~s}^{-1} \mathrm{~Mpc}^{-1}$ \citep{2020_planck} and previous inverse distance ladder measurements from \cite{Aubourg_2015} and \cite{Macaulay_2019}, who found $H_0=67.3\pm1.1\mathrm{~km} \mathrm{~s}^{-1} \mathrm{~Mpc}^{-1}$ and $H_0=67.8\pm1.3\mathrm{~km} \mathrm{~s}^{-1} \mathrm{~Mpc}^{-1}$ respectively (see Fig.~\ref{fig:h0_comparison}). The \cite{Macaulay_2019} $H_0$ measurement was also determined using the $4^{\mathrm{th}}$ order cosmographic model. However, the uncertainties in our measurement are smaller by $\sim50\%$, highlighting the advancements in constraining power from both the \snbao~datasets. Our measurement is inconsistent at $\sim4.6\sigma$ to the SH0ES collaboration measurement of $H_{0}=73.04 \pm 1.04 \mathrm{~km} \mathrm{~s}^{-1} \mathrm{~Mpc}^{-1}$ \citep{SH0ES_2021} determined with the local distance ladder.

We find $M_B=-19.375^{+0.018}_{-0.017}$, which is lower than the value quoted by the SH0ES collaboration of $M_B=-19.253\pm 0.027$, using Cepheid calibrated SNe Ia. We note that our value is significantly lower than the value quoted by \cite{Macaulay_2019} and \cite{Aubourg_2015}, who both find $M_B\approx-19.1$. However, our result is in agreement with various other works who calibrate SN with high redshift data \citep{Camarena_2020, PhysRevD.107.063513, camarena2023tension}.

Given that we have anchored the SNe to the BAO scale, which was in turn calibrated to the CMB values of the sound horizon $r_s$ and drag scale $r_d$, it is not surprising that we find a $H_0$ in agreement with Planck. It is clear from Fig.~\ref{fig:dm_expansion} that although the SNe do prefer a higher $H_0$, once the BAO have set the absolute scale on the vertical axis of $c\ln(1+z)/D_M(z)$ then it is very difficult for the supernovae to change the shape of the expansion history sufficiently to recover a $H_0$ ($y$-intercept of Fig.~\ref{fig:dm_expansion}) above 70~\kmsMpc. 

Here, the supernova data is adding a more detailed expansion history constraint at mid-redshift to low-redshift than the BAO provide. Fig.~\ref{fig:H0_constraint} compares the expansion histories of the $4^{\mathrm{th}}$ order cosmography fit using DESI-BAO alone and that of the fit to the combined \snbao, demonstrating the improvement in constraining power from the combination of data sets. Both expansion histories differ from that of Planck, which assumes Flat-$\Lambda$CDM, but agree with each other.  This is in line with the interesting result that DESI-BAO and DES-SN5YR found consistent non-$\Lambda$CDM expansion histories when fitting for a model that allowed time-varying dark energy \citep[$w_0w_a$CDM, see][]{DES-SN5YR,desicollaboration2024desi}.

In Fig.~\ref{fig:h0_comparison} we compare our result for $H_0$ with other results in the literature. The result we present here is the first inverse distance ladder measurement that has an uncertainty on $H_0$ as small as that from the CMB. However, we note that it does depend on the drag scale $r_d$ calibrated from the CMB so it is not an entirely independent measurement. 

\section{Conclusions}\label{sec:conclusion}
We have used the recent DES supernova data \citep{DES-SN5YR} calibrated by the recent DESI BAO measurements \citep{desicollaboration2024desi} to measure an `inverse distance ladder' estimation of \bestfit. The uncertainties on our $H_0$ measurement are $\sim50\%$ smaller than the inverse distance ladder measurement using the DES 3 year sample \citep{Macaulay_2019} and comparable with the Planck measurement. Interestingly, while our value of $H_0$ agrees with the best-fitting Planck $H_0$ value in Flat-$\Lambda$CDM, the expansion history differs substantially from that of the best fit Planck model (see Figs.~\ref{fig:H0_constraint} \&~\ref{fig:dm_expansion}). This reflects the fact that DES and DESI both prefer a time-varying equation of state of dark energy, $w$, over a cosmological constant.  Alternatively, it remains important to consider that there might be an unexposed systematic error.  This motivates the efforts to continue acquiring more and better data to investigate these cosmological tensions further.

\section*{Acknowledgements}
{\footnotesize 
\textbf{Author Contributions:}
RC contributed to the development of the pipeline, performed the analysis, drafted the manuscript. TMD supervised the project and helped writing. SRH contributed to the code analysis and provided suggestions for the manuscript.

\textit{Construction and validation of the DES-SN5YR Hubble diagram:} 
PA, DB, LG, JL, CL, MSa, DS, PS, MSu. \textit{Contributed to the internal review process:} 
PA, DB, LG, JL, CL, PS, NS, MSu. The remaining authors have made contributions to this paper that include, but are not limited to, the construction of DECam and other aspects of collecting the data; data processing and calibration; developing broadly used methods, codes, and simulations; running the pipelines and validation tests; and promoting the science analysis.

TMD, RC, SH, acknowledge the support of an Australian Research Council Australian Laureate Fellowship (FL180100168) funded by the Australian Government. MV was partly supported by NASA through the NASA Hubble Fellowship grant HST-HF2-51546.001-A awarded by the Space Telescope Science Institute, which is operated by the Association of Universities for Research in Astronomy, Incorporated, under NASA contract NAS5-26555. LG acknowledges financial support from the Spanish Ministerio de Ciencia e Innovaci\'on (MCIN) and the Agencia Estatal de Investigaci\'on (AEI) 10.13039/501100011033 under the PID2020-115253GA-I00 HOSTFLOWS project, from Centro Superior de Investigaciones Cient\'ificas (CSIC) under the PIE project 20215AT016 and the program Unidad de Excelencia Mar\'ia de Maeztu CEX2020-001058-M, and from the Departament de Recerca i Universitats de la Generalitat de Catalunya through the 2021-SGR-01270 grant. 

Funding for the DES Projects has been provided by the U.S. Department of Energy, the U.S. National Science Foundation, the Ministry of Science and Education of Spain, 
the Science and Technology Facilities Council of the United Kingdom, the Higher Education Funding Council for England, the National Center for Supercomputing 
Applications at the University of Illinois at Urbana-Champaign, the Kavli Institute of Cosmological Physics at the University of Chicago, 
the Center for Cosmology and Astro-Particle Physics at the Ohio State University,
the Mitchell Institute for Fundamental Physics and Astronomy at Texas A\&M University, Financiadora de Estudos e Projetos, 
Funda{\c c}{\~a}o Carlos Chagas Filho de Amparo {\`a} Pesquisa do Estado do Rio de Janeiro, Conselho Nacional de Desenvolvimento Cient{\'i}fico e Tecnol{\'o}gico and 
the Minist{\'e}rio da Ci{\^e}ncia, Tecnologia e Inova{\c c}{\~a}o, the Deutsche Forschungsgemeinschaft and the Collaborating Institutions in the Dark Energy Survey. 

The Collaborating Institutions are Argonne National Laboratory, the University of California at Santa Cruz, the University of Cambridge, Centro de Investigaciones Energ{\'e}ticas, 
Medioambientales y Tecnol{\'o}gicas-Madrid, the University of Chicago, University College London, the DES-Brazil Consortium, the University of Edinburgh, 
the Eidgen{\"o}ssische Technische Hochschule (ETH) Z{\"u}rich, 
Fermi National Accelerator Laboratory, the University of Illinois at Urbana-Champaign, the Institut de Ci{\`e}ncies de l'Espai (IEEC/CSIC), 
the Institut de F{\'i}sica d'Altes Energies, Lawrence Berkeley National Laboratory, the Ludwig-Maximilians Universit{\"a}t M{\"u}nchen and the associated Excellence Cluster Universe, 
the University of Michigan, NSF's NOIRLab, the University of Nottingham, The Ohio State University, the University of Pennsylvania, the University of Portsmouth, 
SLAC National Accelerator Laboratory, Stanford University, the University of Sussex, Texas A\&M University, and the OzDES Membership Consortium.

Based in part on observations at Cerro Tololo Inter-American Observatory at NSF's NOIRLab (NOIRLab Prop. ID 2012B-0001; PI: J. Frieman), which is managed by the Association of Universities for Research in Astronomy (AURA) under a cooperative agreement with the National Science Foundation.

The DES data management system is supported by the National Science Foundation under Grant Numbers AST-1138766 and AST-1536171.
The DES participants from Spanish institutions are partially supported by MICINN under grants ESP2017-89838, PGC2018-094773, PGC2018-102021, SEV-2016-0588, SEV-2016-0597, and MDM-2015-0509, some of which include ERDF funds from the European Union. IFAE is partially funded by the CERCA program of the Generalitat de Catalunya.
Research leading to these results has received funding from the European Research
Council under the European Union's Seventh Framework Program (FP7/2007-2013) including ERC grant agreements 240672, 291329, and 306478.
We  acknowledge support from the Brazilian Instituto Nacional de Ci\^encia
e Tecnologia (INCT) do e-Universo (CNPq grant 465376/2014-2).

This manuscript has been authored by Fermi Research Alliance, LLC under Contract No. DE-AC02-07CH11359 with the U.S. Department of Energy, Office of Science, Office of High Energy Physics.
}

\textit{Software}:
{\sc numpy} \citep{numpy}, 
{\sc astropy} \citep{astropy13,astropy18}, 
{\sc matplotlib} \citep{matplotlib}, 
{\sc pandas} \citep{pandas}, 
{\sc scipy} \citep{scipy}, 
{\sc ChainConsumer} \citep{hinton16},
{\sc Dynesty} \citep{dyn3, dyn4, dyn5, dyn1,dyn2}.

\section*{Data Availability}
The {\sc dynesty} chains can be found at \url{https://github.com/RyanCamo/inversedistanceladder/}.


\bibliographystyle{mnras}
\bibliography{mybib} 


\noindent \\ 

\bsp	
\label{lastpage}

\end{document}